# Radar evidence of subglacial liquid water on Mars

**Authors:** R. Orosei[1,2*], S. E. Lauro[3], E. Pettinelli[3], A. Cicchetti[4], M. Coradini[5], B. Cosciotti[3], F. Di Paolo[1], E. Flamini[5], E. Mattei[3], M. Pajola[6,7], F. Soldovieri[8], M. Cartacci[4], F. Cassenti[9], A. Frigeri[4], S. Giuppi[4], R. Martufi[9], A. Masdea[10], G. Mitri[11], C. Nenna[12], R. Noschese[4], M. Restano[13], R. Seu[9]

**Affiliations:**

[1]Istituto Nazionale di Astrofisica, Istituto di Radioastronomia, Via Piero Gobetti 101, 40129 Bologna, Italy.

[2]Università di Bologna, Dipartimento di Scienze Biologiche, Geologiche e Ambientali, Via Zamboni 67, 40126 Bologna, Italy.

[3]Università degli Studi Roma Tre, Dipartimento di Matematica e Fisica, Via della Vasca Navale 84, 00146 Roma, Italy.

[4]Istituto Nazionale di Astrofisica, Istituto di Astrofisica e Planetologia Spaziali, Via del Fosso del Cavaliere 100, 00133 Roma, Italy.

[5]Agenzia Spaziale Italiana, Via del Politecnico, 00133 Roma, Italy.

[6]NASA, Ames Research Center, Moffett Blvd, Mountain View, CA 94035, USA.

[7]Istituto Nazionale di Astrofisica, Osservatorio Astronomico di Padova, Vicolo Osservatorio 5, 35122 Padova, Italy.

[8]Consiglio Nazionale delle Ricerche, Istituto per il Rilevamento Elettromagnetico dell'Ambiente, Via Diocleziano 328, 80124 Napoli, Italy.

[9]Università degli Studi di Roma "La Sapienza", Dipartimento di Ingegneria dell'Informazione, Elettronica e Telecomunicazioni, Via Eudossiana 18, 00184 Roma, Italy.

[10]E.P. Elettronica Progetti s.r.l., Via Traspontina 25, 00040 Ariccia (RM), Italy.

[11]International Research School of Planetary Sciences, Università degli Studi "Gabriele d'Annunzio", Viale Pindaro 42, 65127 Pescara (PE), Italy.

[12]Danfoss Italia, Via Roma 2, 39014 Postal (BZ), Italy.

[13]Serco, c/o ESA Centre for Earth Observation, Largo Galileo Galilei 1, 00044 Frascati (RM), Italy.

*Correspondence to: roberto.orosei@inaf.it.

**Abstract**: The presence of liquid water at the base of the Martian polar caps has long been suspected but not observed. We surveyed the Planum Australe region using the Mars Advanced Radar for Subsurface and Ionosphere Sounding, a low-frequency radar on the Mars Express spacecraft. Radar profiles collected between May 2012 and December 2015, contain evidence of liquid water trapped below the ice of the South Polar Layered Deposits. Anomalously bright subsurface reflections were found within a well-defined, 20km wide zone centered at 193°E, 81°S, surrounded by much less reflective areas. Quantitative analysis of the radar signals shows that this bright feature has high dielectric permittivity >15, matching water-bearing materials. We interpret this feature as a stable body of liquid water on Mars.

**One Sentence Summary:** Strong radar echoes from the bottom of the Martian southern polar deposits are interpreted as being due to the presence of liquid water under 1.5 km of ice.

**Main Text:** The presence of liquid water at the base of the Martian polar caps was first hypothesized over thirty years ago *(1)* and has been inconclusively debated ever since. Radio Echo Sounding (RES) is a suitable technique to resolve this dispute, because low frequency radars have been used extensively and successfully to detect liquid water at the bottom of terrestrial polar ice sheets. An interface between ice and water, or alternatively between ice and water-saturated sediments, produces bright radar reflections *(2,3)*. The Mars Advanced Radar for Subsurface and Ionosphere Sounding (MARSIS) instrument on the Mars Express spacecraft *(4)* is used to perform RES experiments at Mars *(5)*. MARSIS has surveyed the Martian subsurface for more than twelve years in search of evidence for the presence of liquid water *(6)*. Strong basal echoes have been reported in an area close to the thickest part of the South Polar Layered Deposits (SPLD), Mars' southern ice cap *(7)*. These features were interpreted as due to the propagation of the radar signals through a very cold layer of pure water ice having negligible attenuation *(7)*. Anomalously bright reflections were subsequently detected in other areas of the SPLD *(8)*.

On Earth, the interpretation of radar data collected above the polar ice sheets is usually based on the combination of qualitative (the morphology of the bedrock) and quantitative (the reflected radar peak power) analyses *(3, 9)*. The MARSIS design, particularly the very large footprint (~3-5km), does not provide high spatial resolution, strongly limiting its ability to discriminate the presence of subglacial water bodies from the shape of the basal topography *(10)*. Therefore, an unambiguous detection of liquid water at the base of the polar deposit requires a quantitative estimation of the relative dielectric permittivity (hereafter just permittivity) of the basal material, which determines the radar echo strength.

Between 29 May 2012 and 27 December 2015, MARSIS surveyed a 200 km-wide area of Planum Australe, centered at 193°E, 81°S (Fig. 1), which roughly corresponds to a previous study area *(8)*. This area does not exhibit any peculiar characteristics, neither in topographic data from the Mars Orbiter Laser Altimeter (MOLA) (Fig.1A) *(11, 12)* nor in the available orbital imagery (Fig. 1B) *(13)*. It is topographically flat, composed of water ice with 10-20% admixed dust *(14, 15)*, and seasonally covered by a very thin layer of $CO_2$ ice that does not exceed 1 m in thickness *(16, 17)*. In the same location, higher frequency radar observations performed by the Shallow Radar (SHARAD) on the Mars Reconnaissance Orbiter *(18)*, barely revealed internal layering in the SPLD and did not detect any basal echo (Fig. S1), in marked contrast with the North Polar Layer Deposits (NPLD) and other regions of the SPLD *(19)*.

A total of 29 radar profiles were acquired using onboard unprocessed data mode *(5)* by transmitting closely spaced radio pulses centered at either 3 and 4MHz, or at 4 and 5MHz (Table S1). Observations were performed when the spacecraft was on the night side of Mars to minimize ionospheric dispersion of the signal. Figure 2A shows an example of a MARSIS radargram collected in the area, where the sharp surface reflection is followed by several secondary reflections produced by the interfaces between layers within the SPLD. The last of these echoes represents the reflection between the ice-rich SPLD and the underlying material (hereafter basal material). In most of the investigated area the basal reflection is weak and diffuse, but in some locations it is very sharp and has a greater intensity (bright reflections) than the surrounding areas and the surface (Fig. 2B). Where the observations from multiple orbits overlap, the data acquired at the same frequency have consistent values of both surface and subsurface echo power (Fig. S2).

The two-way pulse travel time between the surface and basal echoes can be used to estimate the depth of the subsurface reflector and map the basal topography. Assuming an average signal velocity of 170m/μs within the SPLD, close to that of water ice *(20)*, the depth of the basal reflector is about 1.5km below the surface. The large size of the MARSIS footprint and the diffuse nature of basal echoes outside the bright reflectors prevent a detailed reconstruction of the basal topography, but a regional slope from west to east is recognizable (Fig. 3A). The subsurface area where the bright reflections are concentrated is topographically flat and surrounded by higher ground, except on its eastern side where there is a depression.

The permittivity, which provides constraints on the composition of the basal material, can in principle be retrieved from the power of the reflected signal at the base of the SPLD. Unfortunately, the radiated power of the MARSIS antenna is unknown because it could not be calibrated on the ground (due to the instrument's large dimensions), and thus the intensity of the reflected echoes can only be considered in terms of relative quantities. It is common to normalize the intensity of the subsurface echo to the surface value *(21)*, i.e. to compute the ratio between basal and surface echo power. Such a procedure has the advantage of also compensating for any ionospheric attenuation of the signal. Following this approach, we normalized the subsurface echo power to the median of the surface power computed along each orbit; we found that all normalized profiles at a given frequency yield consistent values of the basal echo power (Fig. S3). Figure 3B shows a regional map of basal echo power after normalization; bright reflections are localized around 193°E, 81°S in all intersecting orbits, outlining a well-defined, 20km wide subsurface anomaly.

To compute the basal permittivity, we also require information about the dielectric properties of the SPLD, which depend on the composition and temperature of the deposits. As the exact ratio between water ice and dust is unknown *(15)* and because the thermal gradient between the surface and the base of the SPLD is poorly constrained *(22)*, we explored the range of plausible values for such parameters and computed the corresponding range of permittivity values. The following general assumptions were made: i) the SPLD is composed of a mixture of water ice and dust in varying proportions (from 2% to 20%); and ii) the temperature profile inside the SPLD is linear, starting from a fixed temperature at the surface (160K) and rising to a variable temperature at the base of the SPLD (range 170 – 270K). Various electromagnetic scenarios were computed *(5)* by considering a plane wave impinging normally onto a three-layer structure: a semi-infinite layer with the permittivity of free space, a homogeneous layer representing the SPLD, and another semi-infinite layer representing the material beneath the SPLD, with variable permittivity values. The output of this computation is an envelope encompassing a family of curves that relate the normalized basal echo power to the permittivity of the basal material (Fig.4A). This envelope is used to determine the distribution of the basal permittivity (inside and outside the bright area) by weighting each admissible value of the permittivity with the values of the probability distribution of the normalized basal echo power (Fig.4B). This procedure generated two distinct distributions of the basal permittivity estimated inside and outside the bright reflection area (Figs. 4C and S4), whose median values (at 3, 4 and 5MHz) are (30±3, 33±1, 22±1) and (9.9±0.5, 7.5±0.1, 6.7±0.1), respectively. The basal permittivity outside the bright area is in the range 4-15, typical for dry terrestrial volcanic rocks. It is also in agreement with previous estimates of 7.5-8.5 for the material at the base of the SPLD *(23)* and with values derived from radar surface echo power for dense dry igneous rocks on the Martian surface at mid-latitudes *(24, 25)*. Conversely, permittivity values as high as those found within the bright area have not previously been observed on Mars. On Earth, values greater than 15 are seldom associated with dry materials *(26)*. RES data collected both in Antarctica *(27)* and in Greenland *(9)* show that a permittivity larger than 15 is indicative of the presence of liquid water below the polar deposits. Based on the evident analogy of the physical phenomena on Earth and Mars, we can infer that the high permittivity values retrieved for the bright area below the SPLD are due to (partially) water-saturated materials and/or layers of liquid water.

We examined other possible explanations for the presence of the radar bright area below the SPLD (supplementary online text). For example, the presence of a $CO_2$ ice layer at the top or the bottom of the SPLD, or a very low temperature of the $H_2O$ ice throughout the SPLD, could enhance basal echo power compared to surface reflections. We reject these explanations (supplementary online text), either because of the very specific and unlikely physical conditions required, or because they do not cause sufficiently strong basal reflections (Figs. S5 & S6). Although the pressure and the temperature at the base of the SPLD would be compatible with the presence of liquid $CO_2$, its relative dielectric permittivity is much lower (about 1.6) *(28)* than that of liquid water (about 80), so does not produce bright reflections.

The discovery of substantial amounts of magnesium, calcium, and sodium perchlorate in the soil of the northern plains of Mars by the Phoenix lander's Wet Chemistry Lab *(29)* supports the presence of liquid water at the base of the polar deposits. Perchlorates can form through different physical/chemical mechanisms *(30, 31)* and were detected in different areas of Mars. It is therefore reasonable to presume that they are also present at the base of the SPLD. Because the temperature at the base of the polar deposits is estimated to be around 205K *(32)* and because perchlorates strongly suppress the freezing point of water (to a minimum of 204K and 198K for magnesium and calcium perchlorates respectively) *(29)*, we therefore find it plausible that a layer of perchlorate brine could be present at the base of the polar deposits. The brine could be mixed with basal soils to form a sludge or could lie on top of the basal material to form localized brine pools *(32)*.

The lack of previous radar detections of subglacial liquid water has been used to support the hypothesis that the polar caps are too thin for basal melting and has led some authors to state that liquid water may be located deeper than previously thought *(e.g., 33)*. The MARSIS data show that liquid water can be stable below the SPLD at relatively shallow depths (about 1.5 km), thus constraining models of Mars' hydrosphere. The limited raw data coverage of the SPLD (a few percent of the area of Planum Australe) and the large size required for a meltwater patch to be detectable by MARSIS (several kilometers in diameter, several tens of centimeters in thickness) limit the possibility to identify small bodies of liquid water or the existence of any hydraulic connection between them. Because of this, there is no reason to conclude that the presence of subsurface water on Mars is limited to a single location.

**Acknowledgments:** We gratefully acknowledge the work of Giovanni Picardi (1936-2015), who served as Principal Investigator of MARSIS. The MARSIS instrument/experiment was funded by the Italian Space Agency and NASA and developed by the University of Rome, Italy, in partnership with NASA's Jet Propulsion Laboratory, Pasadena, Calif. Alenia Spazio (now Thales Alenia Space - Italy) provided the instrument's digital processing system and integrated the parts, and now operates the instrument/experiment. The University of Iowa, Iowa City, built the transmitter for the instrument, JPL built the receiver and Astro Aerospace, Carpinteria, Calif., built the antenna. This research has made use of NASA's Astrophysics Data System. The perceptually-uniform color map broc is used in this study to prevent visual distortion of the data. We thank Marco Mastrogiuseppe and Giuliano Vannaroni for insightful discussions. We are grateful to Stanley E. Beaubien for careful proofreading of the manuscript and improvement of the English language.

**Funding:** This work was supported by the Italian Space Agency (ASI) through contract I/032/12/1.

**Author contributions:** R.O. devised the data calibration method, produced maps of subsurface reflectors, developed the electromagnetic propagation model, co-developed the method for data interpretation and co-wrote the paper. S.E.L. contributed to the development of the electromagnetic propagation model, co-developed the method for data interpretation and co-wrote the paper. E.P. coordinated the writing of the paper, contributed to data analysis interpretation and provided discussion of ideas. A.C. planned and conducted the search for bright subsurface radar reflectors using raw data. M.Co., B.C., F.D.P., E.F., E.M., and M.P. contributed text and figures to the manuscript, and provided discussion of ideas. F.S. contributed to the forward and inverse modeling of the electromagnetic propagation and scattering and provided discussion of ideas. M.Ca., F.C., A.F., S.G., R.M., A.M., G.M., C.N., R.N., M.R. and R.S. contributed to data acquisition and analysis and provided discussion of ideas.

**Competing interests:** There are no competing interests to declare.

**Data and materials availability:** Data reported in this paper, scripts used to model electromagnetic propagation and their output are available through the Zenodo research data repository at the following URL: https://zenodo.org/record/1285179.

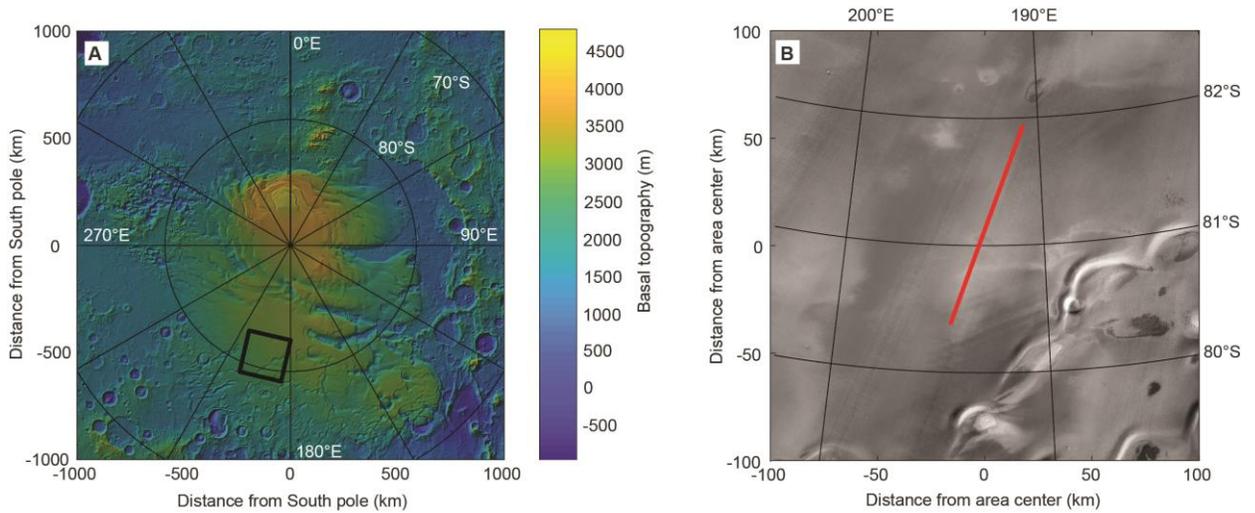

**Fig. 1**. **Maps of the investigated area.** (A) Shaded relief map of Planum Australe, Mars, south of 75°S latitude. The map was produced using the MOLA topographic dataset *(11)*. The black square outlines the study area. (B) mosaic produced using infrared observations by THEMIS (Thermal Emission Imaging System) camera *(13)* corresponding to the black square in panel A. South is up in the image. The red line marks the ground track of orbit 10737, corresponding to the radargram shown in Fig. 2A. The area consists mostly of featureless plains, except for a few Large Asymmetric Polar Scarps (LAPS) near the bottom right of panel B image, which suggests an outward sliding of the polar deposits *(34)*.

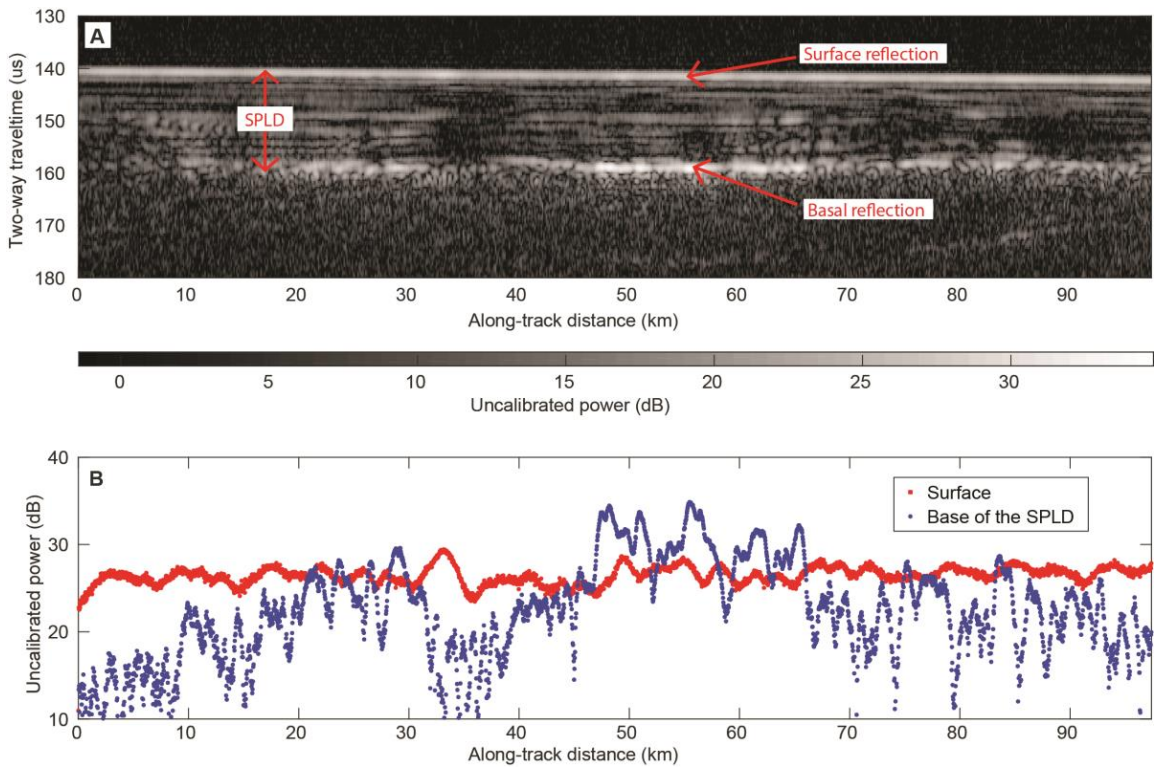

**Fig. 2. Radar data collected by MARSIS.** (A) Radargram for MARSIS orbit 10737, whose ground track is shown in Fig. 1B. A radargram is a bi-dimensional color-coded section made of a sequence of echoes in which the horizontal axis is the distance along the ground track of the spacecraft, the vertical axis represents the two-way travel time of the echo (from a reference altitude of 25km above the reference datum), and brightness is a function of echo power. The continuous bright line in the topmost part of the radargram is the echo from the surface interface, whereas the bottom reflector at about 160μs corresponds to the SPLD/basal material interface. Strong basal reflections can be seen at some locations, where the basal interface is also planar and parallel to the surface. (B) Plot of surface and basal echo power for the radargram in (A). Red dots mark surface echo power values, while blue ones mark subsurface echo power. The horizontal scale is along-track distance, as in (A), while the vertical scale reports uncalibrated power in decibels (dB). The basal echo between 45km and 65km along track is stronger than the surface echo even after attenuation within the SPLD.

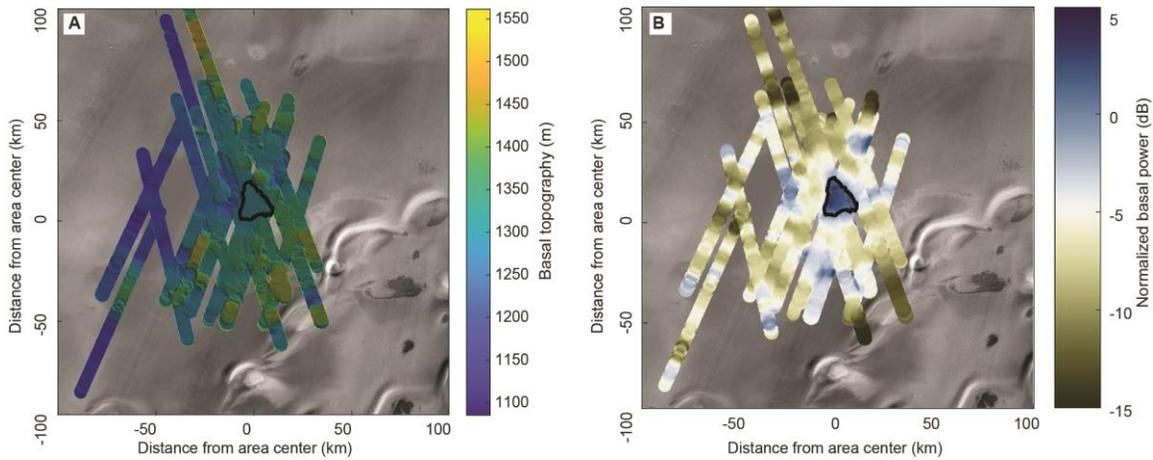

**Fig. 3. Maps of basal topography and reflected echo power.** (A) Color-coded map of the topography at the base of the SPLD computed with respect to the reference datum. The black contour outlines the area in which bright basal reflections are concentrated. (B) Color-coded map of normalized basal echo power at 4MHz. The large blue area (positive values of the normalized basal echo power) outlined in black corresponds to the main bright area, however the map also shows other small bright spots that have a limited number of overlapping profiles. Both panels are superimposed on the infrared image shown in Fig.1B and the value at each point is the median of all radar footprints crossing that point.

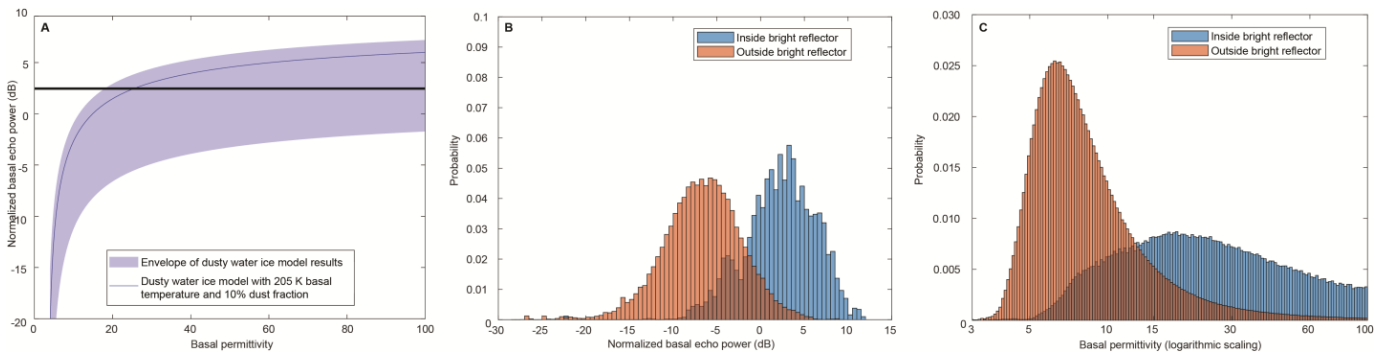

**Fig. 4. Results of the simulation and retrieved permittivities** (A) Output of the electromagnetic simulations computed at 4MHz (see Figs. S4 and S6). The blue shaded area is the envelope of all curves incorporating different amounts of $H_2O$ ice and dust along with various basal temperature for the SPLD. The blue line is the curve for a single model (205K basal temperature and 10% dust content), shown for illustration, and the black horizontal line is the median normalized basal echo power at 4MHz from the observations. (B) Normalized basal echo power distributions inside (blue) and outside (brown) the bright reflection area, indicating two clearly distinct populations of values. These distributions, together with the chart in panel (A), are used to compute the basal permittivity; for example, the intersection between the blue curve and the black line gives a basal permittivity value of 24. (C) Basal permittivity distributions inside (blue) and outside (brown) the bright reflection area. The non-linear relationship between the normalized basal echo power and the permittivity produces an asymmetry (skewness) in the distributions of the values.

**Supplementary Materials:**

Materials and Methods

Supplementary Text

Figures S1-S6

Table S1

References (*35-52*)

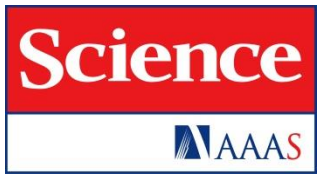

# Supplementary Materials for
## Radar evidence of subglacial liquid water on Mars

R. Orosei, S. E. Lauro, E. Pettinelli, A. Cicchetti, M. Coradini, B. Cosciotti, F. Di Paolo, E. Flamini, E. Mattei, M. Pajola, F. Soldovieri, M. Cartacci, F. Cassenti, A. Frigeri, S. Giuppi, R. Martufi, A. Masdea, G. Mitri, C. Nenna, R. Noschese, M. Restano, R. Seu

correspondence to: roberto.orosei@inaf.it

**This PDF file includes:**

    Materials and Methods
    Supplementary Text
    Figs. S1 to S6
    Table S1
    References (35-52)

**Materials and Methods**

<u>Instrument description</u>

MARSIS is a subsurface radar sounder on the European Space Agency's Mars Express orbiter. It transmits 1 MHz bandwidth pulses centered at 1.8, 3, 4 or 5 MHz, alternating the transmission at two different frequencies. Pulses are 10-W, 250-μs chirped waveforms (linear frequency modulation) transmitted through a 40-m dipole antenna with a repetition frequency of 127.7 Hz *(35)*. The radar collects echoes reflected by the surface and by any other dielectric discontinuity present in the subsurface. Longer wavelengths (lower frequencies) have deeper penetration, but pulse frequency must be above the plasma frequency of the Martian ionosphere to reach the surface.

MARSIS acquires data only when the spacecraft altitude is less than 800-1000 km and operates best in the dark (local night) as echoes suffer far less distortion from the ionosphere.

The radar vertical resolution is approximately 210 m in free space after range compression (i.e., the correlation between transmitted and received waveform) and Hanning windowing (reducing the amplitude of side-lobes caused by range compression) *(36)*. In the subsurface, vertical resolution is improved by a factor equal to the square root of the soil permittivity. Horizontal resolution depends on surface roughness, altitude of the satellite and operating frequency. For most Martian areas, MARSIS lateral resolution is about 10-30 km whereas along track resolution is 5-10 km after Synthetic Aperture Radar (SAR) processing *(35)*, which consists of coherent summation of a batch of consecutive pulses after correction for the vertical motion of the spacecraft.

<u>Operation Mode and Data calibration</u>

Due to spacecraft data transmission rate limitations, MARSIS is typically programmed to perform SAR processing on board *(35)*. Differently, for the present analysis, raw echo data was used. To achieve this, some instrument software parameters had to be modified *(37)*, so that the raw data bypassed the on-board processing and were stored directly in the instrument memory for the subsequent downlink. This new data collection protocol yielded 3200 consecutive echoes, at two different frequencies, over a continuous ground track approximately 100 km long. Data processing on Earth consisted of range compression and geometric calibration to compensate for altitude variations. In our analysis, SAR processing was not performed because of the smoothness of the SPLD in this area, which causes surface echoes to originate solely from the specular direction; in this case, SAR processing would be reduced to a simple moving average of nadir echoes. Moreover, no correction for ionosphere distortion *(38)* was applied to the data.

The radiation pattern of MARSIS antenna could not be characterized before launching, due to its large size, thus preventing retrieval of absolute transmitted power. The only possible form of calibration is the correction of geometric power fall-off due to altitude variations. According to the radar equation *(39)*, radar echo power decreases as the inverse of the fourth power of distance between the antenna and the target.

Because of the surface smoothness in this area, topographic roughness is well below the MARSIS wavelength *(11, 12)* and scattering is almost totally coherent. Under these conditions, the size of the MARSIS footprint is well approximated by the first Fresnel zone. The radius of this zone ranges between 3 and 5 km, depending on satellite altitude (300-800 km) and frequency *(36)*. The power reflected by a flat disk is proportional to the square of its area *(40)* and the area of the Fresnel circle increases linearly with altitude. Substituting these quantities in the radar equation, it is found that the decrease of echo power is inversely proportional to the square of distance. To correct geometric power fall-off due to altitude variations, surface echo power is thus normalized by the squared altitude of the spacecraft.

Data characteristics and normalization

The collected data can be classified into three general categories according to the intensity and variability of the echoes (see Table S1). Almost a third of the radargrams are characterized by a high signal-to-noise ratio and an almost constant surface power along the ground track. Where different orbits overlap, data acquired with the same frequency show consistent values of both surface and subsurface echo power (Fig. S2). The second subset consists of a limited number of radar profiles whose data show sudden drops of surface echo strength, which are not correlated with the noise level nor with subsurface echo power. These variations take place over distances of a few to several kilometers, and do not depend in any obvious way on solar longitude, solar zenith angle or Martian year. Their location and occurrence seem to change with time, as there are instances in which surface echo power decreases seen in one radar profile are not observed in another overlapping one. Their cause remains unclear, but we hypothesize that they result from patches of $CO_2$ ice of variable thickness (on the order of 10 m), affecting MARSIS surface echoes in a way similar to those collected over the south residual cap *(41)*. The remaining profiles (third subset) are characterized by a lower signal-to-noise ratio, presumably due to the ionosphere which, in this case, reduces both surface and subsurface echo power and causes signal distortion *(42)*.

All these data have been used to retrieve the basal permittivity by normalizing the power echoes with respect to the median of the surface power computed along each radar profile. Such a normalization has been used to: i) check the uniformity of the surface (Fig. S3 panels A, C, and E) and ii) calibrate the subsurface (Fig. S3 panels B, D, and F). The use of the median has minimized the effects caused by local surface echo power fluctuations observed in some data, without altering the spatial variation of the basal reflectivity along the profiles.

Electromagnetic propagation model

To simulate MARSIS radar echoes from the surface and the base of the SPLD, a one-dimensional (1-D) electromagnetic plane wave propagation model was used, similar to those presented in *(43)* and *(44)*. The model uses the recursive formula *(45)* to compute the global reflection coefficient of a plane parallel stratigraphy at normal incidence as a function of frequency:

$$R_{i-1}(\omega) = \frac{R_{i-1,i}(\omega) + R_i(\omega) \cdot e^{-2jk_i(\omega)L_i}}{1 + R_{i-1,i}(\omega) \cdot R_i(\omega) e^{-2jk_i(\omega)L_i}} \qquad (S1)$$

where $k_i(\omega)$ is the wave number of the *i*-th layer, $L_i$ the thickness of the *i*-th layer, and $R_{i-1,i}(\omega)$ is the reflection coefficient at the boundary between layer *i-1* and *i,* given by*:*

$$R_{i-1,i}(\omega) = \frac{\sqrt{\varepsilon_{i-1}(\omega)}-\sqrt{\varepsilon_i(\omega)}}{\sqrt{\varepsilon_{i-1}(\omega)}+\sqrt{\varepsilon_i(\omega)}} \tag{S2}$$

where $\varepsilon$ is the layer permittivity.

Planum Australe is modelled as a three-layer structure: a semi-infinite layer with the free space permittivity for the space between the spacecraft and the surface, a homogeneous layer representing the SPLD, and another semi-infinite layer for the material beneath the SPLD. The model computes the echo produced by this stratigraphy, when illuminated by a MARSIS radar pulse under a normal incidence, by iterating equation (S1) for every layer in the model stratigraphy and for every frequency in the MARSIS broadband pulse. Both surface and basal echo power (intensity of the reflected waves) are extracted from the simulated signal and their ratio computed.

Dielectric model of the SPLD

The SPLD are represented as a single homogeneous layer consisting of a mixture of $H_2O$ ice and dust. Internal layering is neglected to simplify computations and reduce the number of model parameters, thus ignoring signal losses due to (possibly multiple) reflections within the SPLD. The effect of $CO_2$ ice is accounted for separately, simulating the radar response of the overall deposit when layers of $CO_2$ ice of variable thickness are located at the top or the base of the SPLD (see Supplementary Text below). The permittivity of water ice is computed according to *(46)*, while that of $CO_2$ ice is taken from *(47)*. Dust within the SPLD is assumed to have a complex permittivity of 8.8+0.017i, typical of basaltic rocks constituting the surface of Mars *(43)*. The permittivity of a mixture of materials is computed by using the Maxwell-Garnett dielectric mixing model *(48)*.

The loss tangent of $H_2O$ ice is strongly dependent on temperature at MARSIS frequencies *(46)*. A value of 160 K is used to represent the mean annual surface temperature of the SPLD *(1)*, while temperature at the bottom of the SPLD is varied between 170 K and 270 K to account for uncertainties in the Martian geothermal flux and basal thermal properties. The temperature profile within the SPLD is assumed to be a linear interpolation between surface and basal temperatures. Dust content of the SPLD is assumed to range between 2% and 20%, based on current estimates *(15)*. The permittivity at the base of the SPLD, being an unknown parameter, is varied between 3, a value typical of dry porous materials, and 100, which is above that of pure liquid water and is indicative of brines *(49)*.

The baseline model of the radar response of the SPLD consists of an $H_2O$ ice layer, as thick as the one observed in radargrams over the bright reflector (Fig. 2), containing variable amounts of dust and having variable basal temperature and permittivity. The model produces a set of curves expressing the relation between the normalized basal echo power and the basal permittivity for this ample parameter space. These are used to determine the distribution of the basal permittivity (inside and outside the bright area), by weighting each admissible value of the permittivity in the chart with the values of the probability distribution of the normalized basal echo power (Fig. 4 and Fig. S4).

**Supplementary Text**

Alternative stratigraphic scenarios for the SPLD

$CO_2$ ice is considered one of the components of the Martian polar caps, and it has been identified in the south residual cap *(41, 50)*. Because it has a lower permittivity compared to the one of water ice *(47)*, $CO_2$ ice could affect basal echo strength, especially in the case of a resonant $CO_2$ ice layer (i.e., that produces constructive or destructive signal interference) either at the top *(41)* or at the base of the SPLD. Occurrence of a $CO_2$ ice layer in the study area could be hypothesized based on the following evidence:

- Data from both Mars Global Surveyor MOLA *(17)* and 2001 Mars Odyssey HEND (High Energy Neutron Detector) *(16)* confirm the presence of a seasonally and locally variable deposit of $CO_2$ ice less than one meter thick over most of Planum Australe;
- Reflection-free zones (RFZs) observed in SHARAD radargrams *(50)* are distinct subsurface volumes within which internal layering is only barely discernible or totally absent, in contrast to that observed in the surrounding and underlying terrains. A RFZ has been identified over the study area (covering in fact a much larger region), which is locally about 170m thick. RFZs possess different radar characteristics in different parts of Planum Australe; only a specific type of RFZ in the residual cap could be identified as consisting of $CO_2$ ice *(50)*.

We are not aware of any direct or indirect observation of layers of $CO_2$ ice close to or at the base of the SPLD, in agreement with theoretical considerations on the instability of $CO_2$ ice at large depths *(51)*.

Modelling the effect of a $CO_2$ layer on radar echo power requires consideration of both its thickness and its depth, which largely increases the model parameter space. The model results have been computed for a $CO_2$ ice layer that is either at the top or at the bottom of the SPLD, which are hypothesized to consist of $H_2O$ ice with a dust content fixed at 10% *(14)*, a basal temperature of 205 K *(32)* and a basal permittivity equal to that of the dust within the SPLD *(43)*. Results related to the CO2 ice layer at the top of the SPLD are shown in Fig. S5 (A and C) and confirm that the seasonal layer of $CO_2$ ice (less than 1 m thick) has no effect on surface reflectivity. For a thickness close to a quarter of the transmitted wavelength, i.e., 10-20 m of $CO_2$ ice, destructive interference between reflections at the top and the bottom of a surface $CO_2$ ice layer would drastically reduce surface reflectivity *(41)*; however, such a large variation in reflectivity is not observed in the data (Fig. S2). Analysis of a similarly resonant, 10-20 m thick $CO_2$ ice layer at the bottom of the SPLD resulted in an increase of the normalized basal echo power by about 3 dB (see Fig. S5B and S5D), well below what is observed in the MARSIS data for the bright reflector area.

For a $CO_2$ ice layer thicker than a few tens of meters, the dominant effect in enhancing subsurface to surface echo power ratio is a weaker attenuation within the $CO_2$ ice layer. The effect of a thick $CO_2$ ice layer covering the entire study area, as in the case of the RFZ mapped by *(50)*, was numerically assessed for a fixed $CO_2$ ice layer thickness of 200 m. The model was run for the same range of SPLD dust content and basal temperatures considered for the $CO_2$-free SPLD model, and results are displayed in Fig. S6. By using these results to derive the value of basal permittivity corresponding to a measured subsurface-to-surface echo power ratio, for all possible combinations of dust content and basal temperature, we still find that estimates within the bright reflector are markedly higher compared to the ones observed elsewhere in the study area. The median values of these permittivity distributions are 14, 16 and 13 (at 3, 4 and 5 MHz) inside the bright reflector and 7, 6 and 6 outside of it. Within the bright reflector, many points have values higher than 15 (more than 50% at 4 MHz), whereas only a few percent of them reach this value outside of bright reflector area (about 4% at 4 MHz).

    A final possibility to explain the occurrence of strong basal echoes without a high basal permittivity is that the SPLD are made of very cold, very pure $H_2O$ ice. Such a possibility seems to be ruled out by estimates of the SPLD density in this area *(15)* and by the layering observed in MARSIS radargrams (Fig. 2), but it was proposed by *(7)* for bright basal reflections observed in other parts of Planum Australe. Results of forward electromagnetic simulations based on this assumption are shown in Fig. S6 as a cyan curve. Even in this case, the median value of the normalized basal echo power at 4 MHz corresponds to a basal permittivity of 17, again above the values typical for dry volcanic rocks.

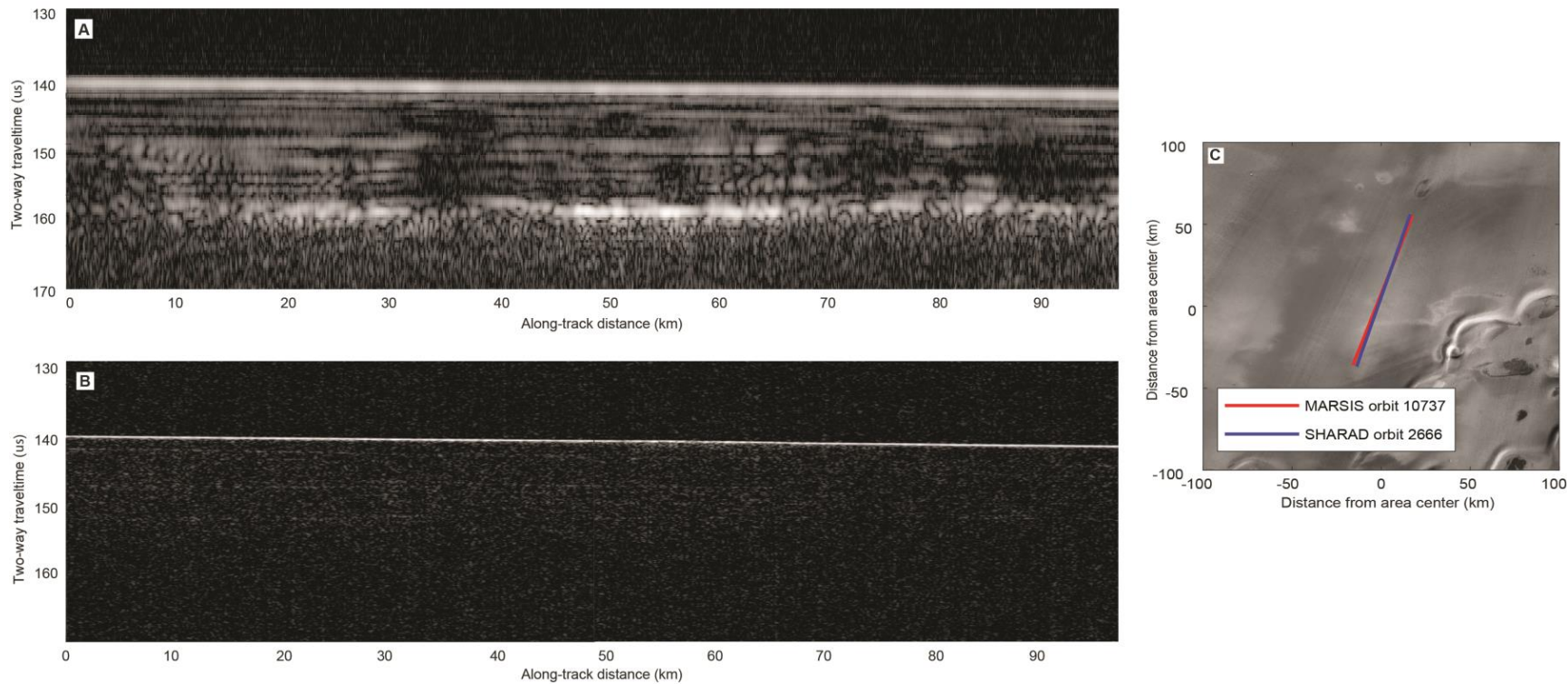

**Fig. S1. Comparison between MARSIS and SHARAD radar data.** (A) MARSIS (orbit 10737) and (B) SHARAD (orbit 2666) radargrams collected in the area with strong basal reflectivity. (C) The ground tracks corresponding to the two radargrams are projected on the same infrared image as Fig. 1B. No basal echo is visible in the SHARAD data, while layering is only faintly discernible amid a diffuse echo attributed to volume scattering in the SPLD between the surface and the basal layer *(52)*, which weakens or masks reflections from its internal layered structure.

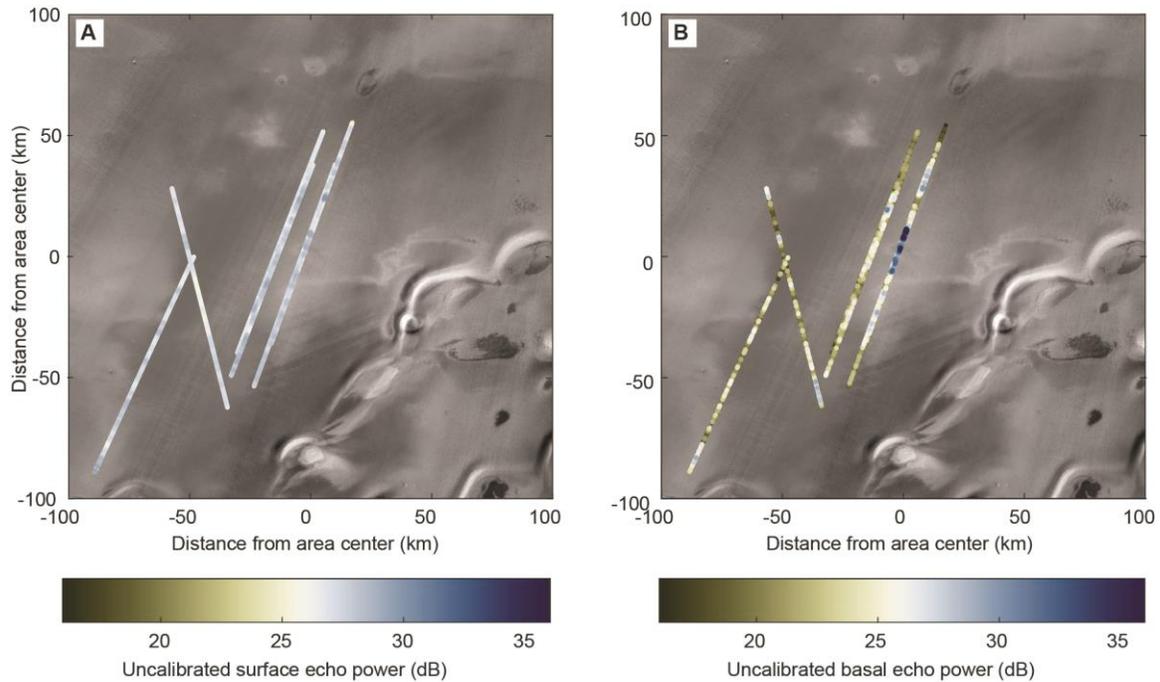

**Fig. S2 Color-coded representation of surface (A) and basal (B) echo power at 4 MHz for MARSIS orbits 10737, 12840, 12847, 12995, 14948 and 14967.** Ground tracks are projected on the same infrared image as Fig. 1B. The data represent the measured echo power after applying only the correction of geometric power fall-off due to altitude variations. The width of the ground tracks has been reduced with respect to the real one to allow the separation of parallel, partially overlapping orbits. Surface echo power fluctuations are limited to a few dB, while there is clustering of strong basal echoes that are consistently observed in different orbits crossing the study area. This indicates that the physical properties of the surface are spatially constant whereas the ones at the base of the SPLD show some lateral variations.

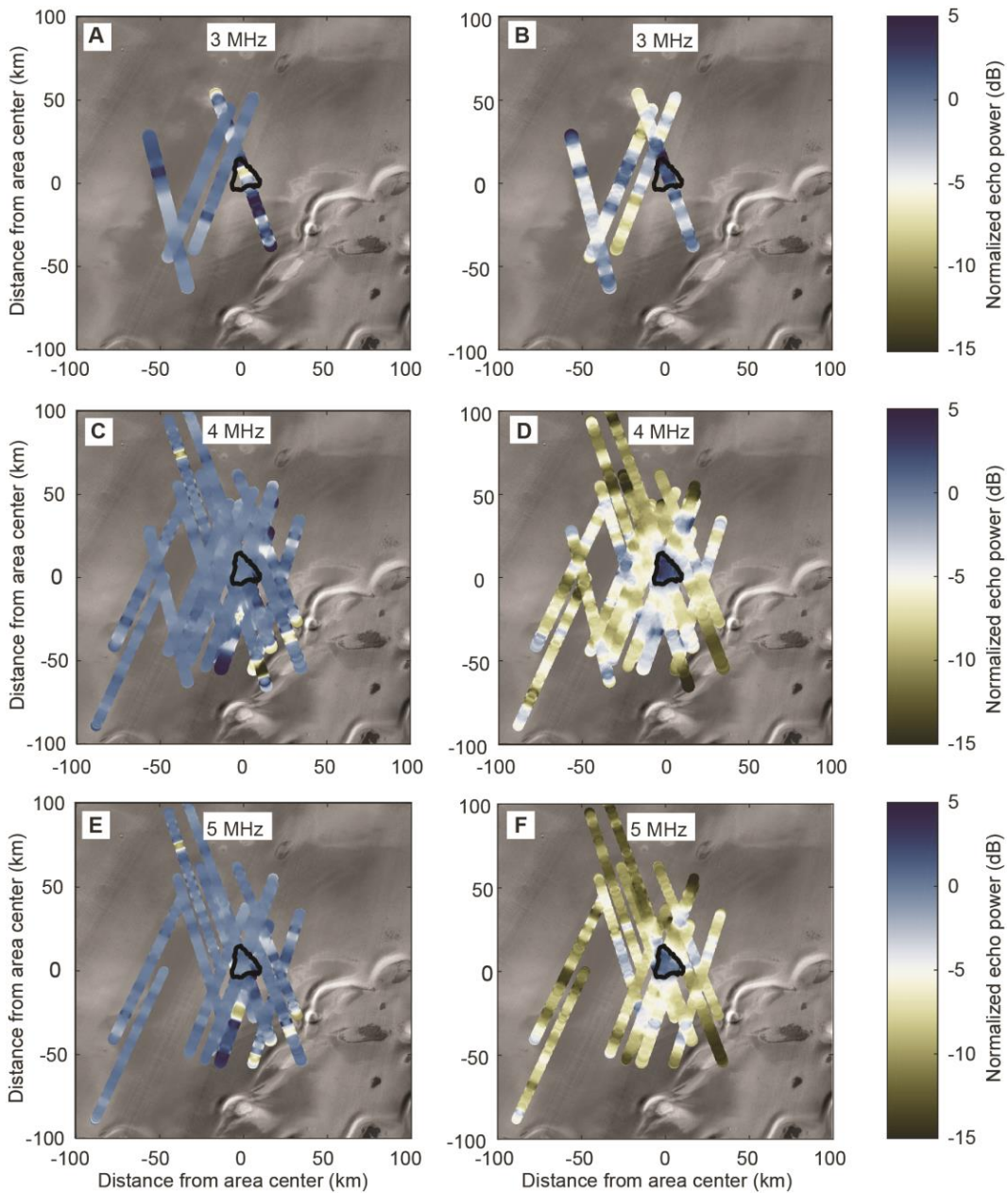

**Fig. S3 Color-coded map of normalized surface (A, C, E) and basal (B, D, F) echo power superimposed on the infrared image in Figure 1B.** Because of the topographic smoothness of the area, the radar footprint size is assumed to correspond to the Fresnel zone of the radar pulse (3-5 km). The mapped value at a given point is the median echo power of all footprints crossing that point. Panels A, C and E show that the surface radar reflectivity is essentially uniform over the area, with only minor, localized and time-varying (seasonal) fluctuations. Panels B, D and F illustrate that the bright basal reflector at the center of the map is visible at all frequencies. The surface reflectivity in the bright area does not exhibit peculiar characteristics compared to the surrounding terrain, except at 3 MHz which we ascribe to limited coverage and low data quality.

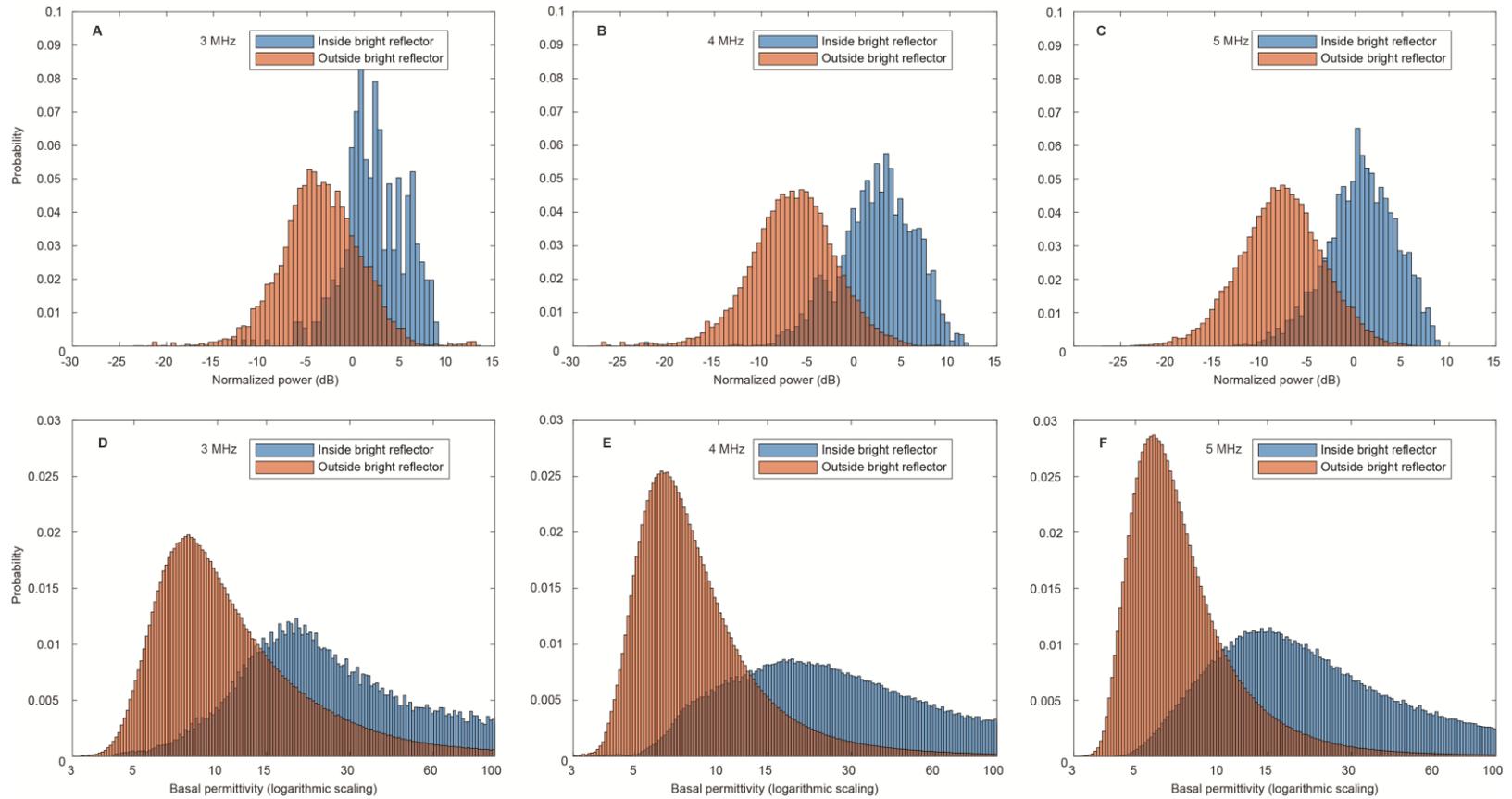

**Fig. S4 Histogram of normalized basal echo power (A-C) and corresponding permittivity values derived from the results of electromagnetic simulations (D-F) at 3, 4 and 5 MHz.** Histograms are computed separately for data points falling inside and outside the bright basal reflector outlined in Fig. 3. The two data sets have clearly distinct statistical properties, and points within the bright reflector produce estimates of basal permittivity larger (median values between 22 and 33) than the values typical of dry volcanic rocks (i.e., 4-15). The long tail of high permittivity values observed in all distributions is caused by the fact that for high values of the basal permittivity, large variations of the permittivity correspond to small variations of basal echo power (Fig. S6). Display is as in Figure 4B-C.

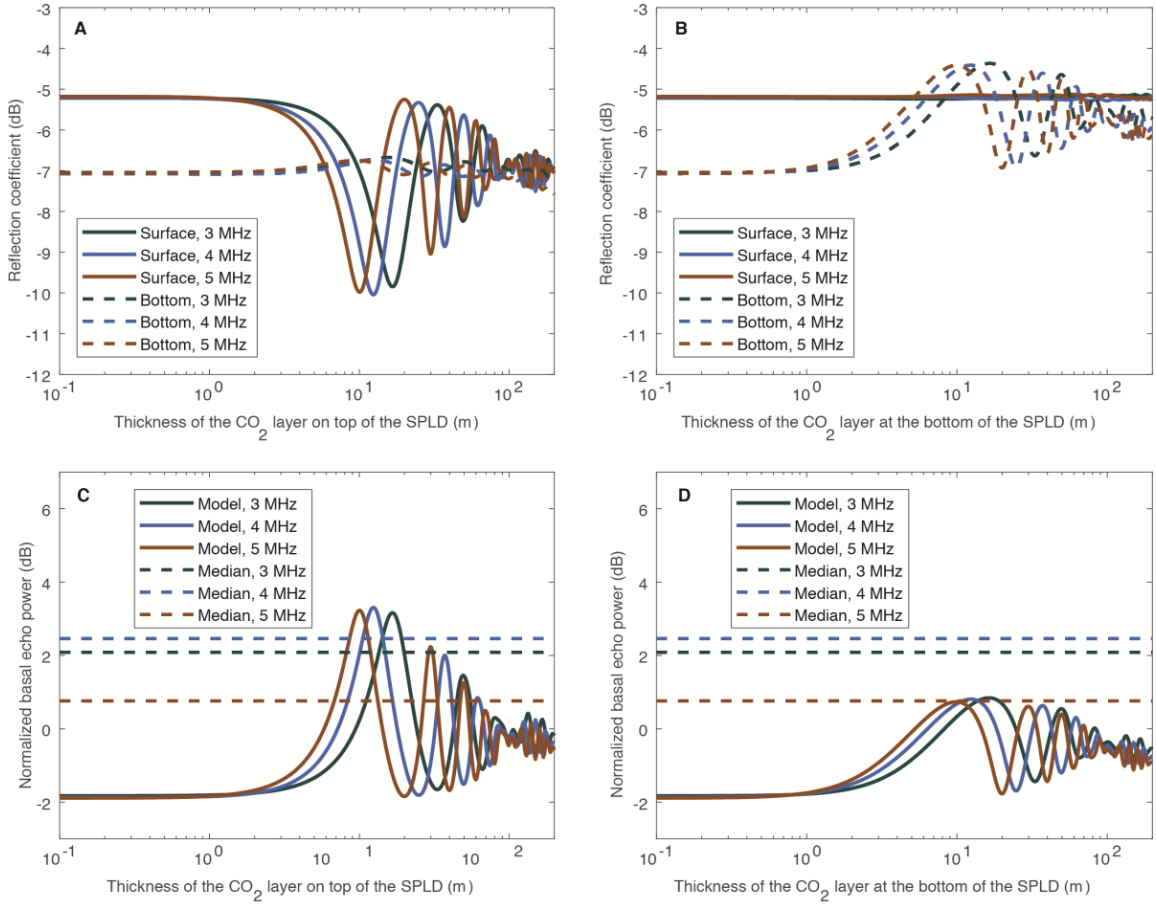

**Fig. S5 Results of electromagnetic propagation simulations for a layer of $CO_2$ ice of variable thickness at the top and the bottom of the SPLD.** Panels A and C correspond to a $CO_2$ ice layer overlying the SPLD, while panels B and D are for a $CO_2$ ice layer beneath it. Panels A and B present values of surface and subsurface reflection coefficients for the three different frequencies employed by MARSIS over the study area. Panels C and D show the corresponding subsurface to surface power ratio compared to the median value observed in MARSIS data inside the bright reflector (dashed lines). The physical properties of the SPLD have been kept fixed (dust content 10%, basal temperature 205 K) for reference. A basal layer of $CO_2$ ice cannot explain the high values of subsurface to surface power ratio observed within the bright basal reflector. A surface $CO_2$ ice layer could cause an enhancement of such a ratio similar to the values observed in the bright reflection area, but it would also result in a decrease of surface reflectivity, which is not observed in the data (cf. Fig. S2).

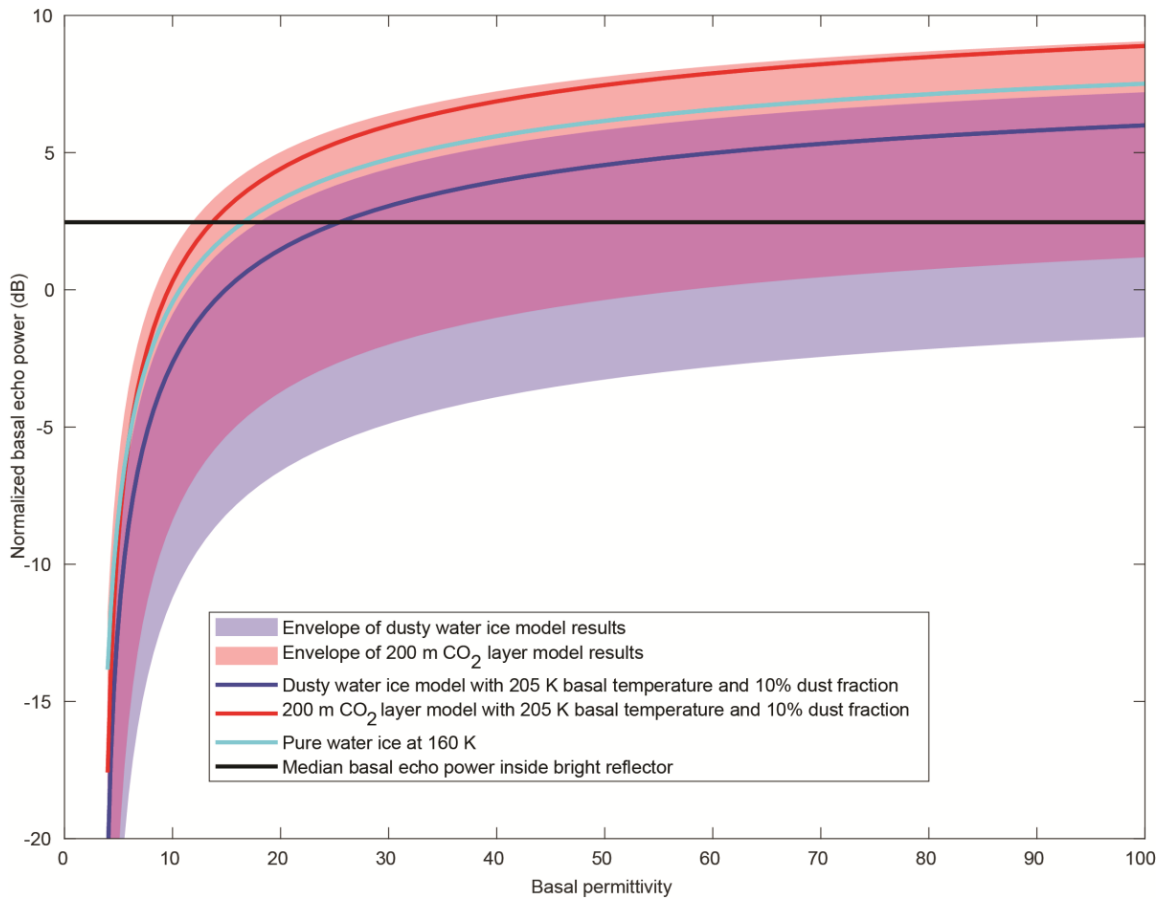

**Fig. S6 Results from numerical simulations of electromagnetic propagation through the SPLD at 4 MHz for the alternative stratigraphic scenarios.** For reference, the blue line is the same as shown in Fig.4. The cyan line corresponds to model results for a pure water ice layer at a uniform temperature of 160 K. The red shaded area encompasses results for the same model SPLD overlain by a 200 m thick layer of $CO_2$, while the red curve refers to model results for a 205 K basal temperature and 10% dust content. The envelopes of results for the two sets of models are mostly overlapping, and they both show that positive values of the normalized basal echo power require a basal permittivity of at least 10.

**Table S1. List of MARSIS profiles over the study area.** Profiles taken simultaneously at two frequencies are listed separately for each frequency. The median surface echo power is computed after correction of geometric power fall-off due to altitude variations. The standard deviation of surface echo power provides an indication of the occurrence of surface reflectivity fluctuations. The notes provide a short assessment of data quality. SNR stands for signal-to-noise ratio.

| Orbit | Frequency (MHz) | Median surface echo power (dB) | Standard deviation of surface echo power (dB) | Notes |
|---|---|---|---|---|
| 10711 | 4 | 29.95 | 3.90 | High SNR, surface echo power fluctuations |
| 10711 | 5 | 22.14 | 4.27 | High SNR, surface echo power fluctuations |
| 10737 | 4 | 27.95 | 1.00 | High SNR |
| 10737 | 5 | 20.20 | 0.86 | High SNR |
| 10961 | 4 | 18.80 | 1.65 | Low SNR, surface echo power fluctuations |
| 10961 | 5 | 10.44 | 1.59 | Low SNR, surface echo power fluctuations |
| 12685 | 4 | 28.98 | 2.03 | High SNR, surface echo power fluctuations |
| 12685 | 5 | 20.63 | 1.68 | High SNR, surface echo power fluctuations |
| 12692 | 4 | N/A | N/A | No signal |
| 12692 | 5 | N/A | N/A | No signal |
| 12759 | 4 | 18.85 | 3.11 | Low SNR, surface echo power fluctuations |
| 12759 | 5 | 13.66 | 2.92 | Low SNR, surface echo power fluctuations |
| 12766 | 4 | 23.14 | 1.30 | Medium SNR, surface echo power fluctuations |
| 12766 | 5 | N/A | N/A | Corrupted data |
| 12814 | 4 | 27.31 | 0.74 | High SNR, surface echo power fluctuations |
| 12814 | 5 | 19.95 | 0.82 | High SNR, surface echo power fluctuations |
| 12840 | 4 | 27.80 | 0.75 | High SNR |
| 12840 | 5 | 20.21 | 0.86 | High SNR |
| 12847 | 4 | 27.70 | 0.89 | High SNR |
| 12847 | 5 | 19.88 | 0.85 | High SNR |
| 12895 | 4 | 24.23 | 3.17 | Medium SNR, reflectivity fluctuations |
| 12895 | 5 | 17.52 | 2.60 | Medium SNR, reflectivity fluctuations |
| 12969 | 4 | 27.80 | 1.19 | High SNR, surface echo power fluctuations |
| 12969 | 5 | 19.88 | 0.98 | High SNR |
| 12995 | 4 | 28.19 | 0.96 | High SNR |
| 12995 | 5 | 20.512 | 0.89 | High SNR |
| 13002 | 4 | 25.74 | 1.20 | Medium SNR |
| 13002 | 5 | 18.73 | 1.07 | Medium SNR |
| 13043 | 4 | 17.87 | 1.96 | Low SNR, surface echo power fluctuations |
| 13043 | 5 | 12.71 | 1.78 | Low SNR, surface echo power fluctuations |
| 13050 | 4 | 20.15 | 0.97 | Low SNR |
| 13050 | 5 | 14.45 | 0.84 | Low SNR |
| 13069 | 4 | 29.24 | 5.57 | High SNR, surface echo power fluctuations |
| 13069 | 5 | 22.42 | 4.58 | High SNR, surface echo power fluctuations |
| 14853 | 4 | 27.96 | 1.40 | High SNR |
| 14853 | 5 | 19.71 | 1.37 | High SNR |
| 14879 | 4 | 25.17 | 0.71 | Medium SNR |
| 14879 | 5 | 17.46 | 0.74 | Medium SNR |
| 14948 | 3 | 23.91 | 1.79 | High SNR |
| 14948 | 4 | 27.13 | 0.74024 | High SNR |
| 14967 | 3 | 24.63 | 1.11 | High SNR |
| 14967 | 4 | 27.18 | 0. 60 | High SNR |
| 15055 | 3 | 23.50 | 0.77 | High SNR |
| 15055 | 4 | 26.46 | 0.49 | High SNR |
| 15110 | 3 | 22.34 | 4.66 | Medium SNR, surface echo power fluctuations |
| 15110 | 4 | 26.62 | 3.79 | High SNR, surface echo power fluctuations |
| 15136 | 3 | N/A | N/A | Distorted signal |
| 15136 | 4 | 31.54 | 3.92 | High SNR, surface echo power fluctuations |
| 15198 | 3 | N/A | N/A | Distorted signal |
| 15198 | 4 | 28.23 | 0.84 | High SNR |